\documentclass[preprint,psfig]{aastex}

\def\d{{\bf d}}
\def\rn{r_{\nu}}
\def\k{{\bf k}}
\def\kp{\k_{\perp}}
\def\kn{k_{\parallel}}
\def\n{{\bf n}}
\def\u{{\bf U}}
\def\up{{\bf U^{'}}}
\def\th{{\vec{ \theta}}}
\def\x{{\bf x}}
\def\v{{\bf v}}
\def\HI{{\rm HI}}

\shorttitle{HI Fluctuations at Large Redshifts}
\shortauthors{Bharadwaj and Sethi}
\begin{document}
  \title{HI Fluctuations at Large Redshifts:
    I--Visibility correlation}
     \author{Somnath Bharadwaj}
\affil{Department of Physics and Meteorology \& Center for Theoretical
Studies,\\
I.I.T. Kharagpur, 721 302, India}
\email{somnath@phy.iitkgp.ernet.in}
\and
\author{Shiv K. Sethi} 
\affil{Harish-Chandra  Research Institute, Chhatnag Road, Jhusi, Allahabad
211 019, India}  
\email{sethi@mri.ernet.in}
\begin{abstract}
  We investigate the possibility of probing the large scale structure
  in the universe at large redshifts by studying fluctuations in the
  redshifted $1420\, \rm MHz$ emission from the neutral hydrogen (HI)
  at early epochs.  The neutral hydrogen content of the universe is known from
  absorption studies for $z \lesssim 4.5$. The HI distribution is
  expected to be inhomogeneous in the gravitational instability
  picture and this inhomogeneity leads to anisotropy in the redshifted
  HI emission. The best hope of detecting this anisotropy is by using
  a large low-frequency interferometric instrument like the Giant
  Meter-Wave Radio Telescope (GMRT). We calculate the
  visibility  correlation function $\langle V_\nu ( \u) V_{\nu
    '} (\u)\rangle$ at two frequencies $\nu$ and $\nu'$
  of the redshifted HI
  emission for an interferometric observation. In particular we give
  numerical results for the two GMRT channels  centered around $\nu =
  325 \, \rm MHz$ and $\nu = 610 \, \rm MHz$ from density
  inhomogeneity and peculiar velocity of the HI distribution.
  The visibility correlation is $\simeq 10^{-10}\hbox{--} 10^{-9} \,
  \rm Jy^2$.  We calculate the signal-to-noise  for detecting  the correlation
  signal in the presence of system  noise and show that the GMRT might
  detect the signal for integration times $\simeq 100 \, \rm hrs$. We
  argue that the measurement of visibility correlation allows
  optimal use of  the uncorrelated
  nature of the system noise across baselines and frequency channels. 
  \end{abstract}

\section{Introduction}
Various observations  indicate that around $90 \%$ of the HI mass 
in the redshift range $2$ to $3.5$ is in clouds which have  HI column  
densities greater than $2 \times 10^{20} {\rm atoms/cm^{2}}$ (Peroux
{\it et al.} 2001, Storrie-Lombardi, McMahon, Irwin 1996, Lanzetta, Wolfe,
\& Turnshek 1995). These  high column density clouds are responsible for
the damped Lyman-$\alpha$ absorption lines  observed  along lines  of
sight to quasars.
 The flux of HI emission from individual clouds ($\lesssim 10 \mu {\rm
Jy}$) is too weak to be detected by existing radio telescopes unless
the image of the cloud is   significantly magnified  by an intervening
cluster gravitational lens (Saini, Bharadwaj and Sethi,
2001). Although we may not be able to detect  
individual clouds, the redshifted HI emission from the distribution of
clouds will appear as  background radiation in   low
frequency radio observations. Bharadwaj, Nath and Sethi (2001;
hereafter referred to as BNS) have  
 used existing estimates of the HI density at $z \simeq 3$ to  infer the
 mean  brightness temperature of $\simeq 1 \, {\rm mK}$
 at $\nu \simeq 320 \, {\rm MHz}$
for this radiation. The fluctuations in the brightness temperature  of
this radiation arise from fluctuations  in the HI number density and
from peculiar velocities.
As shown in BNS,  the cross-correlation between the temperature
fluctuations across different frequencies and  different lines of
sight is related to  the  two-point correlation function (or
equivalently the power spectrum) of  density perturbations at the
redshift where the radiation originates.    
The possibility of measuring  this provides a   new method for
studying  large scale 
structures at high redshifts. Estimates indicate the expected values
of the cross-correlations in the brightness temperature to 
vary from $10^{-7} \, \rm K^2$ to $10^{-8} \, \rm K^2$  
over intervals corresponding to spatial scales from $10 \, {\rm Mpc}$
to $40 \,  {\rm Mpc}$   for some of the currently-favoured
cosmological models. Estimates  of the different contributions to the
flux expected in a pixel of a 
radio image   show the contribution from galactic and extragalactic
sources and  the system noise to be substantially higher than the
contribution from the HI radiation. The task of devising a strategy
for extracting the signal from the various foregrounds and noise  in
which it is buried is a problem which has still to be solved.  A possible 
strategy based on the very distinct spectral properties of the
foregrounds as against the HI emission is discussed in BNS. 

An alternate strategy for using the HI emission from high redshifts
to study large scale structures  has been discussed by many authors
(Subramanian \& Padmanabhan 1993, Kumar, Padmanabhan \& Subramanian,
Bagla, Nath, \& Padmanabhan, Bagla 1998).
This is based on detecting the HI emission from
individual protoclusters at high redshifts. 
There have also been observational efforts in
this direction (see Subrahmanyan \& Anantharamaiah 1990 and 
reference therein). No detections have been made till date. 
This strategy suffers from the disadvantage that 
the protoclusters correspond to  very large overdense regions which 
are very rare events. 
Protoclusters with flux in  the range $3$ to $5 \,  {\rm 
mJy}$ are predicted to occur with abundances in the  range $10^{-8} -
10^{-7} \, {\rm Mpc}^{-3}$ in the CDM model  (Subramanian and
Padmanabhan, 1993). In the statistical approach proposed in BNS
fluctuations of all magnitude in the HI distribution contribute to the
signal.   The statistical approach allows optimum use of the signal 
present in all the pixels of  the  images made at different
frequencies across the bandwidth of a typical radio observation.
In this paper we take up various issues related to  the
statistical approach originally proposed in BNS. 

The main focus of this paper  is the choice of an appropriate 
statistical estimator to quantify the properties of the signal
and the system noise. 
 The statistical estimator proposed in BNS is 
 the cross-correlation between the temperature fluctuations along
different lines of sight in  radio map made at different frequencies.  
While this quantity is conceptually very simple, complications arise
when it is applied to images produced by radio interferometry, as is  the
case with most low frequency radio telescopes. Such observations
measure  the coherence between the signals arriving at any two  antennas, 
a quantity known as the  visibility $V(\u)$. This is measured  for
all pairs of antennas in the  interferometric array. Here  $\u$
refers to the  vector joining a pair of antennas, measured in units of 
$\lambda$,  projected   on  the 
plane perpendicular to the direction which is being imaged.   
 
The image is produced by a Fourier transform of
the visibility
\begin{equation}
I_{\nu}(\th)=\int V(\u) \exp[2 \pi i \,  \th \cdot \u ] \, d^2 \u 
\label{eq:in1}
\end{equation}
where $\th$ refers to different positions in the small patch of the
sky  which is being imaged, and $I_{\nu}(\th)$ is the specific
intensity.    The contribution of system noise to the signal from each
antenna is independent, and the  visibilities measured by each   pair
of antennas are uncorrelated.  The noise in the pixels of a radio image
constructed from the visibilities  is  not independent.
The correlations in the noise in the pixels depend on the detailed
distribution of the different separations (baselines) $\u$ for which
the visibility has  been measured. Any strategy based on the
statistical analysis of radio images  will be faced with the problem
of distinguishing the correlations in fluctuations of the HI emission
from the correlations in the noise. This complication can be avoided
by  dealing directly with the visibilities. In this paper we
investigate the possibility of detecting the fluctuations in the HI
emission using a statistical estimator  constructed directly from the
visibilities  measured in a radio interferometric observation, without
making an image.      

We next present a brief outline of this paper.  In section 2 we
calculate the relation between the statistical properties of the HI
fluctuations  and  the visibilities produced by these fluctuations. In
Section 3 we present numerical estimates of these quantites for some of the
currently favoured cosmological models for the system parameters of
GMRT (Swarup {\it et al.} 1990). In Section 4,  we calculate 
statistical properties of the visibility correlations  arising from the system
noise. In Section 5   we
present the conclusions and discuss possible directions for future work. 

\section{From Density Fluctuations to Visibility}
We consider radio-interferometric observations of a
small patch of the sky whose  center is  in the direction of the unit 
vector $\n$ (Figure 1). A small patch of the sky may be treated as a
plane,  and the angle  $\th$  which refers to different directions in
the sky (Figure 1) may be treated as a two dimensional vector.    
Observations at a frequency $\nu$ would measure the  HI
emission from a redshift  $z=(1420 \, {\rm MHz}/\nu)-1$ or equivalently a
comoving distance $\rn$. 
The specific intensity of the  redshifted HI emission  arriving from
any direction  $\th$ may be decomposed into two 
parts   
\begin{equation}
I_{\nu}(\th)=\bar{I}_{\nu} + \Delta I_{\nu}(\th) \, , \label{eq:a1}
\end{equation}
where $\bar{I}_{\nu}$ and $\Delta I_{\nu}(\n)$ are  the isotropic and
the fluctuating components of the specific intensity. The 
isotropic component $\bar{I}_{\nu}$ is related to $\bar{n}_{\HI}(z)$, 
the mean comoving  number density of HI atoms in the excited  state
at  a redshift $z$  and we have  
\begin{equation}
\bar{I}_{\nu}=\frac{A_{21}\, h_P \, c \, \bar{n}_{\HI}(z)}{4 \pi H(z)}
\label{eq:a2}
\end{equation}
where $A_{21}$ is the Einstein coefficient for the HI hyperfine
transition, $h_P$ the Plank constant, $c$ the speed of light  and
$H(z)$ the Hubble parameter.  This is a slightly rearranged version of
Equation (11) of BNS. 

The fluctuations in  the specific intensity  $\Delta I_{\nu}(\th)$
arise  from  fluctuations in the HI number density 
$\Delta n_{\HI}(\x)$ and  the peculiar velocity  $\v(\x)$,   where
$\x$ refers to the comoving position $\x = \rn (\n + \th)$  
The details of the calculation relating these quantities
are  presented in BNS, and we use a slightly rearranged version 
of  Equation (12) of BNS  
\begin{equation}
\Delta I_{\nu}(\th)= \bar{I}_{\nu}  \left[ \frac{ \Delta n_{\HI}(\x)}{
\bar{n}_{\HI}}  + \frac{(\n \cdot \bf{\nabla}) (\n \cdot \v(\x))}{a H}
\right]  \label{eq:a3}
\end{equation}
where $a$ is the scale factor.   It should be noted that all the quantities 
in the right hand side of equation (\ref{eq:a3}) should be
evaluated at the epoch when the radiation was emitted.   

In this paper we wish to  calculate  the  contribution from the
redshifted HI emission to the visibilities $V_{\nu}(\u)$ that would be
measured  in radio-interferometric observations.    The relation
between the specific intensity and the  visibilities  is  
\begin{equation}
V_{\nu}(\u)= \int d^2 \theta  A(\th) \, \Delta  I_{\nu}(\th) \, e^{- i
2 \pi \u \cdot \th} \, .
\label{eq:a4}
\end{equation}
Only the fluctuating part of the specific intensity 
contributes to the visibility, and we have dropped the isotropic component
from eq. (\ref{eq:a4}). Here $A(\th)$ is the beam pattern of the
individual antennas in the array (primary beam). 
We use equations (\ref{eq:a3}) and (\ref{eq:a4}) to relate the
visibilities to the fluctuations in the HI distribution.  

It is convenient to work with  $\Delta_{\HI}(\k)$, the Fourier
transform of  the density contrast of the HI number density
$\Delta n_{\HI}(\x)/\bar{n}_{\HI}$. We assume that on sufficiently
large scales $\Delta_{\HI}(\k)$ can be related to  $\Delta(\k)$, the
density contrast of the underlying dark matter distribution, through
a linear bias parameter $b$ i.e. $\Delta_{\HI}(\k)=b_{\HI}
\Delta(\k)$.    We also assume that the scales we are dealing with are
sufficiently large that we can apply linear theory of density
perturbations (Peebles 1980) to relate the peculiar velocities to the
fluctuations in the dark matter distribution,  
$\v(\k)=(-i a H f(\Omega_m) \k/k) \Delta(\k)$,  where  
$f(\Omega_m)\approx
\Omega_m^{0.6}+\frac{1}{70}[1-\frac{1}{2}\Omega_m(1+\Omega_m)]$ 
in a spatially flat universe (Lahav {\it et al.} 1991). These assumptions
allow us to express the fluctuations in the specific intensity as 
\begin{equation}
\Delta I_{\nu}(\th) = \bar{I}_{\nu} \int \frac{d^3 k}{(2 \pi)^3} 
\left[1 +  \frac{\beta \kn^2}{k^2} \right] 
\Delta_{\HI}(\k) \, e^{i \rn (\kn + \kp \cdot \th)}
\label{eq:a5}
\end{equation}
where we have decomposed the wave vector $\k$ into two parts $\k=\kn
\n + \kp$ where $\kn \n$ refers to the component of the  Fourier mode
$\k$ along the line of sight to the center of the patch of sky being 
observed, and  $\kp$ refers to the component  of $\k$ in the plane of
the sky. We use equation (\ref{eq:a5}) in equation (\ref{eq:a4}) to
express the visibility $V_{\nu}(\u)$ in terms of $\Delta(\k)$. This
allows us to carry out the  integral over $\th$ which gives us  
\begin{eqnarray}
 V_{\nu}(u,v) =    \bar{I}_{\nu}  \int \frac{d^3 k}{(2 \pi)^3}  
\Delta_{\HI}(\k) \left[1 +  \frac{\beta \kn^2}{k^2} 
\right] e^{i \rn \kn}  \,  a(\u-\frac{\kp \rn}{2 \pi}) 
\label{eq:a6}
\end{eqnarray} 
where $a(\u)$ is the Fourier transform of $A(\th)$ the primary beam, 
\begin{equation} 
a(\u)=\int \, d^2 \th  \,   A(\th) \, e^{-i 2 \pi \u \cdot \th}  \,.
\label{eq:a7}
\end{equation}
For a Gaussian primary beam pattern
$A(\theta)=e^{-\theta^2/\theta_0^2}$.  the Fourier transform also is a 
Gaussian and we have 
\begin{equation}
a(\u)=\pi \theta_0^2 \exp{\left[-\pi^2 \theta^2_0 U^2 \right]}
\label{eq:a8}
\end{equation}
which we use in the rest of this paper. 

Equation (\ref{eq:a6}) relates the contribution to the visibilities
from  fluctuations in the HI number density. These fluctuations are
assumed to be a Gaussian random field, or equivalently  the different 
modes $\Delta_{\HI}(\k)$ have independent, random phases. This allows
us to predict all the statistical properties of $\Delta_{\HI}(\k)$ in
terms of the power spectrum of the fluctuations in the HI distribution
$P_{\HI}(\k)$ which is defined as   $\langle \Delta^{*}_{\HI}(\k)
\Delta_{\HI}(\k^{'}) \rangle = (2 \pi)^3  \delta^3(\k-\k^{'})
P_{\HI}(\k)$ where $\langle \rangle$ denotes  ensemble average. We use
this to calculate the correlation between the 
visibilities at different baselines $\u$ and $\up$ and two different
frequencies $\nu$ and $\nu+ \Delta \nu $. Here we assume that the
bandwidth  over which the observations are being carried out is small
compared to  the central frequency {\it i.e.} $\Delta \nu \ll \nu$
and $r_{\nu +\Delta \nu}=r_{\nu} + r^{'}_{\nu} \Delta \nu$ where
$r^{'}_{\nu}=\frac{d r_{\nu}}{d \nu}$. Using these inputs to calculate
the  visibility  correlation function we obtain 
\begin{eqnarray}
\langle V_{\nu}(\u) V^{*}_{\nu+\Delta \nu}(\up) \rangle &=&
\left[ \bar{I}_{\nu} \theta^2_0 \pi \right]^2 \int  \frac{d^3 k}{(2
\pi)^3} P_{\HI}(k) e^{i \kn  r^{'}_{\nu} \Delta \nu} \left[1+ \beta
\frac{\kn^2}{k^2} \right]^2  \times \nonumber \\
&\times&  \exp \left[ - \frac{(\kp-2 \pi \u/\rn)^2 +
(\kp-2 \pi \u^{'}/\rn)^2}{(2/\rn \theta_0)^2} \right]
\label{eq:a9}
\end{eqnarray}

 The assumption of linear bias allows us to relate  
$P_{\HI}(\k)$ to $P(\k)$ the power spectrum of density fluctuations in
the dark matter distribution  through  the linear bias parameter
$P_{\HI}(\k)=b^2 P(\k)$. We use this in later sections to obtain
numerical estimates for different cosmological models. 

We next turn our attention to a qualitative analysis of equation
(\ref{eq:a9}) to determine the nature and extent of the correlations
between the visibilities measured at different baselines. This is
largely governed by term $ \exp \left[ - \frac{(\kp-2 \pi \u/\rn)^2 +
(\kp-2 \pi \u^{'}/\rn)^2}{(2/\rn \theta_0)^2} \right]$ which arises
because  the observations have a limited sky coverage determined by
the primary beam pattern. This term is very small for 
all values of $\kp$ unless $\mid \u-\u^{'}\mid < 1/\theta_0$.  
The parameter $\theta_0$ is related to the
the FWHM of the primary beam and $\theta_0 \approx 0.6 \, \theta_{{\rm
FWHM}}$, which allows us to relate $\theta_0$ to $D$ the diameter of
the individual antennas as $\theta_0 \approx \lambda/D$. We also
express   $\u$ and $\u^{'}$ in terms of $\d$ and $\d^{'}$,
the physical separations between the pairs of antennas   as $\u=
\d/\lambda$  and $\u^{'}= \d^{'}/\lambda^{'}$ It should be noted that
here and throughout  we assume that $\lambda^{'}=\lambda (1-\Delta
\nu/\nu)$ with $\Delta \nu << \nu$.  Using these we see that the
condition for the  visibilities to be correlated can be expressed as
$\mid \d - \d^{'}\mid < D$. This implies that the visibilities
measured by a  pair of antennas separated by the displacement $\d$
will be correlated to the visibilities measured by another pair
separated by a displacement $\d^{'}$ only if difference in the two
displacements $\d$ and $\d^{'}$ is less than the antenna diameter.  
The consequences of this for a typical antenna configuration are 
\begin{itemize}
\item[(a)] The visibilities measured at various frequencies by the
same pair  of antennas  are correlated. 
\item[(b)]  The visibilities measured by different pairs of  antennas are
uncorrelated.  
\end{itemize}
For the rest of the paper we shall consider only the correlation
between the visibilities measured at various frequencies by the same
pair of antennas. We  use the notation 
$\langle V_{\nu}(\u) V_{\nu + \Delta \nu}^{*}(\u) \rangle $ to denote
the correlation between the visibilities measured at two different
frequencies by the pair of antennas at a physical separation $\d=c \u
/\nu$. The fact that this physical separation $\d$ will correspond to a
different baseline $\u^{'}=(\nu + \Delta \nu) \d /c$  at the frequency
$\nu + \Delta \nu$ is ignored throughout as $\Delta \nu \ll \nu$.  
Equation (\ref{eq:a9}) can now be used to obtain 
\begin{eqnarray}
\langle V_{\nu}(\u) V^{*}_{\nu+\Delta \nu}(\u) \rangle &=&
\left[ \bar{I}_{\nu} \theta^2_0 \pi \right]^2 \int  \frac{d^3 k}{(2
\pi)^3} P_{\HI}(k) e^{i \kn  r^{'}_{\nu} \Delta \nu} \left[1+ \beta
\frac{\kn^2}{k^2} \right]^2  \times \nonumber \\
&\times&  \exp \left[ - \frac{(\kp-2 \pi \u/\rn)^2 }{2 (1/\rn
\theta_0)^2} \right]  
\label{eq:a10}
\end{eqnarray}
The role of the Gaussian in equation (\ref{eq:a10}) arising from the
primary beam pattern is to ensure that most of the contribution is
from Fourier modes for which $\kp \approx (2 \pi/\rn) \u$. Equation
(\ref{eq:a10}) is further simplified if we approximate the Gaussian
with a Dirac Delta function 
\begin{equation}  
\exp \left[ - \frac{(\kp-2 \pi \u/\rn)^2 }{2 (1/\rn
\theta_0)^2} \right]  \approx \frac{2 \pi}{\rn^2 \theta_0^2} \delta^2
\left( \kp- \frac{2 \pi}{\rn} \u \right) 
\label{eq:a11}
\end{equation}
whereby only Fourier modes for which $\kp = (2 \pi/\rn) \u$
contribute. This allows us to do two of the integrals in equation
(\ref{eq:a10}) giving us 
\begin{eqnarray}
\langle V_{\nu}(\u) V^{*}_{\nu+\Delta \nu}(\u) \rangle &=&
\frac{\left[ \bar{I}_{\nu} \theta^2_0 \right]^2}{2} \int_0^{\infty} d \kn
\frac{ P_{\HI}(k)}{\rn^2 \theta_o^2} 
 \cos(\kn  r^{'}_{\nu} \Delta \nu) \times \nonumber \\
&\times& \left[1+ \beta \frac{\kn^2}{k^2} \right]^2 \,\,  {\rm with} \,\,
k=\sqrt{\kn^2+(2 \pi /\rn)^2 U^2} 
\label{eq:a12}
\end{eqnarray}
Equations (\ref{eq:a10}) and  (\ref{eq:a12}) represent the main
results of this section. They relate the correlation in the
visibilities to the power spectrum of fluctuations in the HI
distribution. The visibility correlations at any baseline $U$
are seen to be sensitive to Fourier modes $\ge 2 \pi U/\rn $. It
comes from the fact that each visibility measurement is sensitive to
one Fourier mode in the plane of the sky, which arises from the
projection of three-dimensional Fourier modes  making different angles
with the plane of the sky. The typical length scales $\simeq \pi/k$
that contribute to the measurement for any baseline $U$ are
$\lesssim 30 h^{-1} \, {\rm Mpc} (100/U)$.

In the next section we use these equations to make
predictions for the visibility  correlations expected in
 the currently favoured cosmological models and we discuss the
possibility of observing these. 
\section{Results}
Equation (\ref{eq:a2}) can be calculated to give:
\begin{equation}
\bar{I}_{\nu}=\frac{5.4  \, h \, {\rm Jy}}{{\rm degree^2}} 
\Omega_{gas}(z) \left[ \Omega_{m0} (1+z)^3 + \Omega_{\Lambda 0} \right]^{1/2}
(1+z)^{-3} 
\label{eq:b2}
\end{equation}
for a spatially flat cosmological model. Here $(1+z) = 1420/\nu$,
$\nu$ being the observed frequency.  We  use
$\Omega_{gas}(z)=10^{-3}$ as a fiducial value throughout for $z \ge 1$
(Peroux {\it et al.} 2001). We give results for the currently-favoured
cosmological model: spatially-flat with  $\Omega_{\Lambda 0} =0 .7$
and $ \Omega_{m0} = 0.3$ (Perlmutter {\it et al.} 1999, de Bernardis
{\it et al.} 2000). We use $h = 0.7$ whenever quoting a numerical
value (Freedman {\it et al.} 2001). 

For GMRT $\theta_{\rm FWHM} = 1.8^{\circ}\times (325 \, {\rm
  MHz}/\nu)$. We plot the visibility  correlation function
in Figures~2 and~3 for the  GMRT
channels centered around $\nu = 325 \, \rm MHz$ and $\nu = 610 \rm
MHz$ for COBE-normalized power spectrum (Bunn \& White 1996) and $b
=1$. GMRT has a total bandwidth of $16 \, \rm MHz$ at these  frequencies in 128
channels. The visibility correlation function  shown in the figures is
averaged over one of the these channels with $\Delta \nu = 1.25 \, \rm
kHz$. The correlation function is not very sensitive to the width of
the channel so long as the channel width $\la \hbox{a few} \, \rm kHz$.

For $\nu = \nu'$, the signal ($\sqrt{\langle V_\nu(\u) V_{\nu'}(\u)
  \rangle}$)  is  $10\hbox{--}50 \, \rm \mu Jy$
for baselines $|\u| \simeq 100\hbox{--}1000$.
GMRT has 15 antennas in
a central array within a radius of $\simeq 1 \, \rm km$.  Antenna
pairs formed from these antennas will be most sensitive to the
signal. 

For $\nu \ne
\nu'$, the correlation signal is seen to dip sharply as frequency
separation is increased. The signal is anti-correlated
and drops below $\simeq 1 \, \rm
\mu Jy$ for $\nu' - \nu \ge  2 \rm \, MHz$. For $\nu' - \nu \le 0.5
\rm \, MHz$ the signal is $5\hbox{--}30 \, \rm \mu Jy$ for baselines
$\lesssim 500$.

In Figure~2 and~3 we assume that the HI distribution follows the
underlying dark matter distribution, .i.e. $b =1$. However this may not be
true; observed structures at any  redshift are expected to be  biased
with respect to the underlying mass distribution (see e.g. Bardeen
{\it et al.} 1986). This bias is expected to be higher at larger
redshifts and the observed
strong clustering  of high redshift galaxies is at least
partly owing to this fact (Steidel {\it et al.} 1998). In this paper
we adopt a simple model of bias and assume it to be linear and
independent of the Fourier mode. Though the
exact dependence of the HI signal on the bias is complicated
(Eq.~(\ref{eq:a12}), the signal scales roughly linearly with
bias. This means that for a moderately biased HI distribution $b \le 2$
the signal could be higher by a factor of two.

\section{\bf Noise}

For each visibility measurement  in the UV plane, the contribution
comes from both the signal from HI
fluctuations $S_\nu$, the detector noise $N_\nu$, and various
galactic and extragalactic foregrounds. We consider here the
contribution from only the HI signal and the noise.
The visibility measurement gives:
\begin{equation}
  V_\nu(\u)  = S_\nu(\u) 
  + N_\nu(\u)
  \label{eq:vcor}
\end{equation}
Both $S$ and $N$ are complex quantities with both
real and imaginary parts distributed as Gaussian random variables (see
e.g. Crane \& Napier 1989 for properties of noise). The signal is
a Gaussian random field because it is  linear in  density
perturbation $\Delta (\vec k)$ which is expected to be a Gaussian
random field for large scales (small $k$) (see e.g. Peacock 1999,
Bardeen {\it et al.} 1986). The signal and noise are uncorrelated with
each other. The
reality condition of  the surface brightness (Eq.~\ref{eq:in1}) and
the noise in the real space implies $S(-\u) =
S^*(\u)$ and $N(-\u) = N^*(\u)$
Our aim is to construct  bilinear combinations like
$V_\nu(\u) V_{\nu'}^*(\u)$ and to  detect $S_\nu(\u)
S_{\nu'}^*(\u)$.  $\nu$ and $\nu'$ will in
general be different.  The average
signal $\langle S_\nu(\u) S_{\nu'}^*(\u) \rangle$ is calculated in the
previous section. The average noise correlation, for $\nu = \nu'$:
\begin{equation}
  \langle N_\nu(\u) N_\nu^*(\u) \rangle = \left[{T_{\rm sys}\over K \sqrt{\Delta
        \nu \Delta t } }\right]^2
  \label{eq:nnoi}
\end{equation}
Here $T_{\rm sys}$ is the system temperature, $\Delta \nu$ is the
bandwidth, $K$ is the antenna gain,  and $\Delta t$ is the
time of integration for one visibility
measurement.  For $\nu \ne \nu'$ the noise correlation vanishes as the
noise in different frequency channels is uncorrelated. 

{\it Estimator of the signal}:  From the measured visibility it is
possible to write several estimators of the signal. The simplest such
estimator is:
\begin{equation}
  \hat S = V_\nu(\u) V_{\nu'}^*(\u) - \langle N_\nu(\u) N_\nu^*(\u) \rangle 
  \label{eq:esti}
\end{equation}
This estimator is clearly unbiased, i.e.
$$\langle \hat S \rangle =
\langle S_\nu(\u) S_{\nu'}^*(\u) \rangle. $$
The quantity of
interest to us is the variance of the estimated signal: $\sigma(\hat S)^2 =
\langle \hat S^2 \rangle - \langle \hat S \rangle^2$. This quantity is
calculated
to be (see Appendix A for a derivation):
\begin{equation}
  \sigma^2(\hat S) \simeq  {q \over n}\langle N_\nu(\u) N_\nu^*(\u) \rangle^2
  \label{eq:varesti}
  \end{equation}
  $ q= 2$ for $\nu = \nu'$ and $q = 1$ for $\nu \ne \nu'$.
  $n$ is the total number of visibility measurements.
  The signal-to-noise for the
detection of the HI fluctuation signal is $\langle \hat S
\rangle/\sigma(\hat S)$.

The value of $n$ for a given $|\u|$ in general  depends on the
antenna positions, frequency coverage,
and the position of the source in the sky.
To calculate the value of $n$ we first consider the case 
when $\nu = \nu'$.  

{\it Case I-- $\nu = \nu'$}: To get a simple estimate,
assume that a pair of antennas
describes circular tracks  in the UV  plane with radius $|\u|$
and that these tracks do not overlap
with the tracks of other antenna pairs, then $n = T/\Delta t$ for
each frequency channel, where
$T$ is the total time of observation. (The actual observation is
more complicated because the antenna tracks cross in the UV
plane.) For this case, Eq.~(\ref{eq:varesti}) gives, using
Eq.~(\ref{eq:nnoi}):
\begin{equation}
   \sigma^2(\hat S) = \left[{2T_{\rm sys}\over K\sqrt{\Delta
         \nu \, T } }\right]^2
   \label{eq:noivis}
\end{equation}
Before proceeding further it is useful to calculate the expected noise
for the GMRT. We shall take the
fiducial observing frequency to be $\simeq 320 \, \rm MHz$. At this
frequency the  GMRT system
temperature  $T_s \simeq 110 \, \rm K$. GMRT
has a total bandwidth of $16 \, \rm MHz$ at this frequency in 128
channels.  To calculate the quantity in Eq.~(\ref{eq:nnoi}) we take
one channel ($\Delta \nu = 125 \, \rm kHz$) and an integration time
of $\Delta t = 30 \, \rm sec$  for one 'instantaneous' measurement for
a given baseline. The antenna gain at this frequency $K = 0.32 \,\rm K
\, Jy^{-1}$. Using this Eq.~(\ref{eq:nnoi}) gives the noise
correlation to be  $\simeq 175 \, \rm
mJy$. As this is much larger than the expected signal calculated
in the last section, we are justified in neglecting the signal term in
Eq.~(\ref{eq:app4}) in the appendix.
For total time of integration $T = 10 \, \rm hrs$,
Eq.~(\ref{eq:noivis}) gives $\sigma(\hat S) \simeq 5 \, \rm
mJy$.

As each frequency channel gives a realization of the signal, the
noise can be reduced further by using all the frequency channels.
This gives $n = N_{\rm chan} \times T/\Delta t$ with
$N_{\rm chan} = 128$ for GMRT; Eq.~(\ref{eq:noivis})
gives $\sigma(\hat S) \simeq 0.45 \, \rm
mJy$. Even though we made a few simplifying assumptions in calculating
the noise, this is the typical value obtainable in a real
experiment.
The expected noise is much larger than the expected signal from the
HI fluctuations and therefore an experiment like GMRT cannot detect
the HI fluctuation for a given $|\u|$ or by using a single pair
of antennas in a reasonable amount of integration time.

To reduce the noise further
one must consider averaging the signal over more than one pair of
antennas. One such estimator is the map RMS which uses information
from all possible baselines. The total number of 'instantaneous'
baselines for an experiment with $N$ antennas is $N(N-1)/2$. This
gives $n = N_{\rm chan} N(N-1)/2 \times T/\Delta t$ and 
gives a further decrease of a factor $\simeq N/\sqrt{2}$ in the
sensitivity. In the previous section we showed that much of the
contribution to signal comes from baselines $\le 1000 \lambda$. GMRT
has 15 antennas in the central array within a radius of $\simeq 1 \, \rm
km$. Much of the contribution to the signal will come from these
antenna pairs.

Taking $N =15$ in the calculation of noise sensitivity, we get
$\sigma(\hat S) \simeq 40  \, \rm \mu Jy$ for 10 hours of
integration. The average signal (averaged over 15 antennas of the
central GMRT array) is $\simeq 20 \hbox{--}40 \, \rm \mu Jy$. 
This means that a few sigma detection of the signal might be feasible in
integration time $\le 100 \, \rm hrs$ using the central array of GMRT. 

{\it Case II-- $\nu \ne \nu'$}:  In this case, $\langle N_\nu(\u)
N_\nu^*(\u) \rangle = 0$ and the variance of the signal estimator
is smaller by a factor of 2  (Eq.~(\ref{eq:varesti})). The rest
of the calculations proceeds similar to the first case. The  
number of distinct pairs for two different frequencies will depend on
the separation of the frequencies.
However, one must also take into account the line width of
the  damped Lyman-$\alpha$ clouds which is $\simeq 200 \, \rm km \,
sec^{-1}$ (Prochaska \& Wolfe 1998).
The line width of each GMRT channel is $\simeq 120 \, \rm
km \, sec^{-1}$. This means that one damped Lyman-$\alpha$ cloud will spill
over in many frequency  channels, thereby creating a correlation in
the signal for nearby channels.  This correlation must be accounted for
before the HI fluctuation signal can be extracted. It is hard to do it
analytically and this issue will be addressed in future using
simulations of the HI signal. However it is possible to get the
typical noise sensitivity for this measurement. 

The total number of frequency channel pairs is $N_{\rm chan} (N_{\rm chan}
-1)/2$. Each pair gives a different realization of noise.
If we average the signal over all baselines and all
frequency pairs, we get $n = N(N-1)/2 N_{\rm chan} (N_{\rm
  chan}-1)/2$. This is larger than the maximum value of $n$ in the
first case by a factor of $\simeq N_{\rm chan}/2$. However it would be
meaningful to average over all channel-pairs if  the signal is significant
for all such cross-correlations.
As seen in the previous section, the correlation
between different channels falls rapidly for separation $\ge 1 \, \rm
MHz$, therefore the number of useful channels pairs is less than
the maximum possible. The value of $n$ however is still likely to be
more than in the previous case. For example if we average the signal
over all the frequency channels with $\nu' - \nu \le 0.5 \, \rm MHz$,
the expected signal is $\simeq 10\hbox{--}20 \, \rm \mu Jy$. The
number of frequency pairs are $\simeq 5 N_{\rm chan}$ in this case.
This gives $\sigma(\hat S) \simeq 15  \, \rm \mu Jy$ for ten hours of
integration.
From this discussion we can conclude that it might also be possible to
extract this signal for integration time $\le 100 \, \rm hrs$ using
GMRT.

The noise in detecting the HI signal $\sigma(\hat S)$ is comparable to
the sensitivity for detecting continuum sources. This is so even
though the HI clouds emit line radiation. Therefore the method of
observing  fluctuations in HI radiation makes more optimal use of all
the  frequency width available in the experiment. The individual
clouds are very faint (flux $\la 10 \, \rm \mu Jy$ (Saini {\it et al.}
2001)) and cannot be
detected using GMRT because the line sensitivity $\simeq 50 \, \rm \mu
Jy$ for 100~hrs of integration needed to detect the HI fluctuations. 

\section{Conclusions and Discussion}

Our main results are:
\begin{itemize}
\item[1.] The correlation in measured visibilities owing to  density
  inhomogeneities and peculiar velocities of the 
  HI distribution at high redshifts can be related to the power spectrum of the HI
  distribution (Eq.~(\ref{eq:a12})). The visibility correlation for
  any baseline $U$ is sensitive to Fourier modes $\ge 2 \pi U/\rn $.
  This  means that the typical length scales probed for any baseline
  are $ \lesssim 30 h^{-1} \, {\rm Mpc} (100 /U)$.
  \item[2.] The signal is strongest for baselines $\lesssim 1000$
    and for $\nu = \nu'$, i.e. on the same two-dimesional map,
    the correlation is between $2 \times
    10^{-9} \, \rm Jy^2$  and $10^{-10} \, \rm Jy^2$.
  \item[3.]   For  $\nu \ne \nu'$, i.e., cross-correlation signal,  the
  correlation signal is $ 10^{-9} \hbox{--}  10^{-11} \, \rm Jy^2$
  for baselines $|\u| \lesssim 500 $ for $\nu' - \nu \lesssim
  0.5 \, \rm  MHz$. The correlation is negative for most baselines
  for $\nu' - \nu \gtrsim 2 \, \rm MHz$ and falls below $ 10^{-12}  \,
  \rm Jy^2$.
  \item[4.] GMRT might detect  these signals for integration times
    $\lesssim 100 \, \rm hrs$. We argue that measuring visibility
    correlations in the presence of system noise makes optimal use
    of  the fact that the noise is uncorrelated across baselines and
    frequency channels. The  error for these measurements is
    comparable to and can even be smaller than the continuum
    sensitivity of the instrument.
    \end{itemize}

The signal and noise analyses given in this paper are for the system
parameters of the  currently-operational GMRT.
However it can be easily extended to future
telescopes like Square Kilometer Array (SKA) \footnote{see {\tt
    http://www.nfra.nl/skai/}} and Low Frequency Array (LOFAR)
\footnote{see {\tt http://www.astron.nl/lofar}}. 
Our  analysis can also be extended to higher redshifts ($z \simeq 5$)
as    the HI content of the universe at these redshifts is
beginning to be known (see e.g. Peroux {\it et al.} 2001).

In this paper we neglected two other contributions to the 
visibility correlations: galactic and extra-galactic
foregrounds and the Poisson fluctuations owing to point-like nature of
HI clouds. The galactic foregrounds are expected to be dominated by
the fluctuations in the synchrotron radiation from the
Galaxy. The only existing all-sky map at low radio frequencies  is the
$408 \, \rm MHz$ Haslam map (Haslam {\it et al.} 1982). This map has an angular
resolution of $\simeq 1^\circ$ and therefore cannot give much
information on the angular scales of interest to us. The
extra-galactic foregrounds get most of its contribution from
the radio point-sources. Little is know about the radio point sources
at sensitivity levels ($\lesssim 100 \, \rm \mu Jy$) and frequencies
of relevance in this paper. However it seems likely that these
foregrounds will dominate the HI signal (BNS 2001), and a possible
strategy to remove foregrounds was discussed in BNS (2001). This issue
will be discussed in  a later paper by using simulations of the HI signal and
the foregrounds. 

The HI at large redshift is locked up in discrete clouds. This will
give rise to visibility correlations even in the absence of
gravitational instability. This signal also depends on the
mass function of the HI clouds (Saini {\it et al.} 2001) and therefore
the detection of this signal can give
important clue about  how the HI at high redshift is distributed.
We shall attempt to estimate this signal from simulation of the high
redshift HI in a later publication.

\section*{Acknowledgements}
The authors would like to thank Jayaram Chengalur
for many useful discussions.  We are thankful to late
K. R. Anantharamiah for helpful comments on the noise properties of
interferometric experiments.

\section*{Appendix A}

The HI fluctuation signal and the noise satisfy the following
conditions:
\begin{eqnarray}
  \langle S_\nu (\u) S_{\nu'}(\u') \rangle & = & \langle S_\nu (\u) S_{\nu'}(\u) \rangle \delta_D(\u - \u') \\
  \langle N_\nu (\u) N_{\nu'}(\u') \rangle & = & \langle N_\nu (\u) N_\nu(\u) \rangle \delta_D(\u - \u') \delta_D(\nu -\nu')
\end{eqnarray}
Note that the signal is correlated across frequency channels while the
noise is not. Let us assume that  any interferometric experiment
makes $n$ measurements of the visibility for given $|\u |$ and the
quantities (like visibility correlation function)
are estimated  by averaging over these measurement. Using
this it is seen that the signal (for $\nu = \nu'$) and the noise can be
treated as  $n$ uncorrelated, random variables 
  with the same mean and variance. In such a case, the estimated
  average equals the average of any of the random numbers and the
  variance of the estimated 'signal' is $1/n$ times the variance of
  any of the random variable (see e.g. Papoulis 1965; this result is
  independent of the probability distribution functions of the
  individual variables). Of particular interest to us is the
  variance of the estimator in Eq.~(\ref{eq:esti}) for a given $|\vec
  u|$. The variance is  estimated from $n$ realizations of the
  random variable $\hat S$. It  is given by:
  \begin{equation}
    \sigma^2(\hat S) = {\sigma^2 \over n}
    \label{eq:app1p}
    \end{equation}
Here $\sigma^2 =\langle \hat S^2 \rangle - \langle \hat S \rangle^2$
is the variance of  any realization of $\hat S$.

It is given by, using the definitions of $V$ and
$N$ from Eq.~(\ref{eq:vcor}):
\begin{equation}
  \sigma^2  
   =   \langle V_\nu 
  V_{\nu'} V_\nu V_{\nu'} \rangle + \langle N_\nu N_{\nu'} \rangle^2 - 2
  \langle N_\nu N_{\nu'} \rangle \langle V_\nu V_{\nu'} \rangle -\langle
  S_\nu S_{\nu'} \rangle^2 
  \label{eq:app2}
\end{equation}
To simplify this expression further we use the
fact that for a Gaussian random process, the expectation
value of four random numbers is given by
$$
\langle x_1 x_2 x_3 x_4 \rangle = \langle x_1 x_2 \rangle \langle x_3 x_4 \rangle + \langle
x_1 x_3 \rangle \langle x_2 x_4 \rangle + 
\langle x_2 x_4 \rangle \langle x_1 x_3 \rangle
$$
We first consider the case when $\nu = \nu'$. Eq.~(\ref{eq:app2}) then
reduces to:
\begin{equation}
\sigma^2 = 3 \langle V_\nu V_{\nu'} \rangle^2 + \langle N_\nu N_{\nu'} \rangle^2 - 2
  \langle N_\nu N_{\nu'} \rangle \langle V_\nu V_{\nu'} \rangle -\langle
  S_\nu S_{\nu'} \rangle^2
\end{equation}
Again using Eq.~(\ref{eq:vcor}) for the definition of $\langle V_\nu
V_\nu' \rangle$, this simplifies to:
\begin{equation}
  \sigma^2 = 2 \times \left (\langle  S_\nu S_\nu \rangle + \langle
    N_\nu N_\nu \rangle \right )^2
  \label{eq:app3}
  \end{equation}
  The case when $\nu = \nu'$ is slightly more complicated. Making a
  simplifying assumption that the signal contribution can be dropped
  while calculating the four-point functions in Eq.~(\ref{eq:app2})
  (for justification see the text), we get:
  \begin{equation}
    \sigma^2 \simeq  \langle N_\nu N_\nu \rangle^2
    \label{eq:app4}
  \end{equation}
  Eqs.~(\ref{eq:app3}) and~(\ref{eq:app4}) along with
  Eq.~(\ref{eq:app1p})
  gives Eq.~(\ref{eq:varesti}).

\newpage

\begin{figure}
\figurenum{1}
\plotone{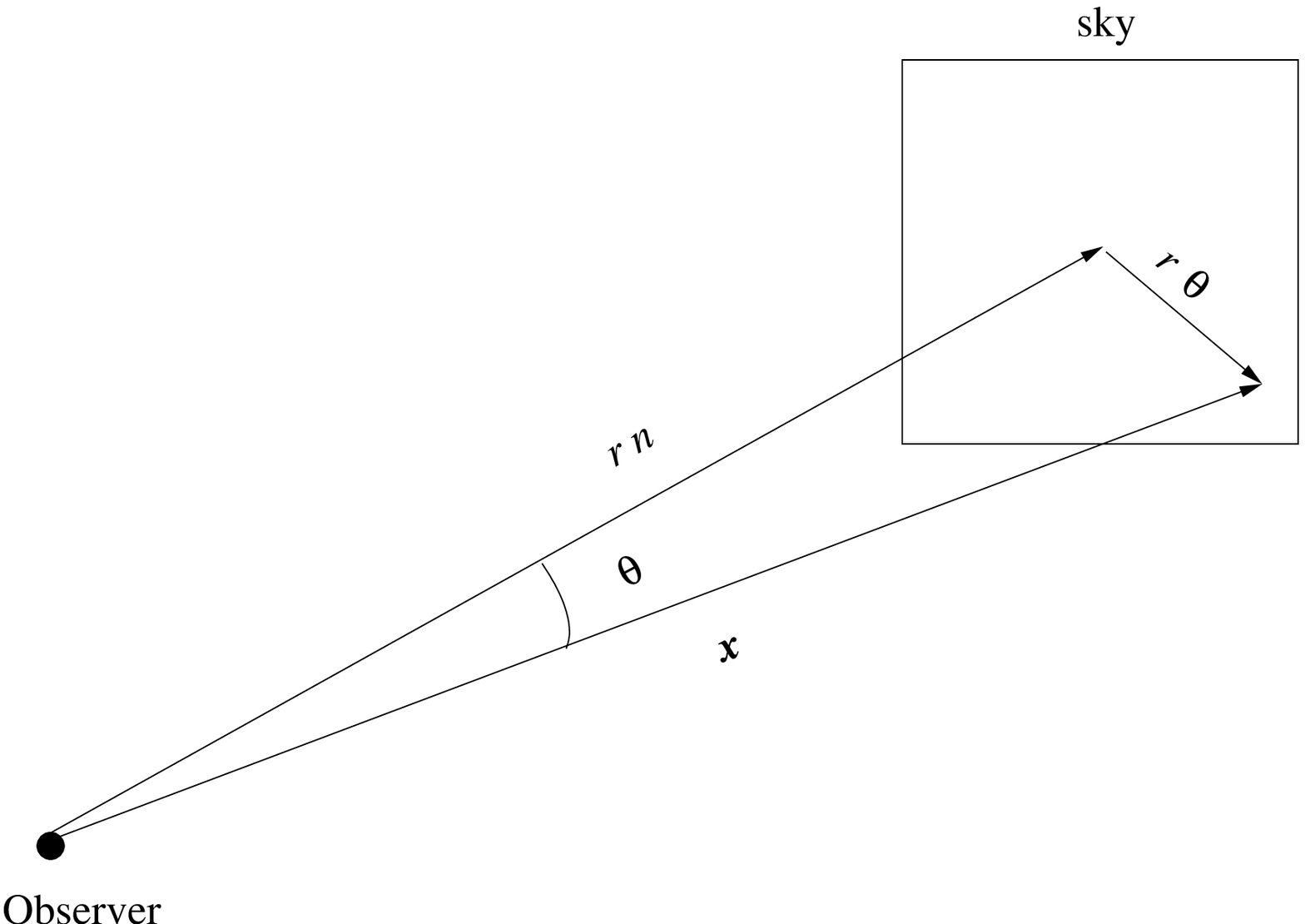}
\caption{The geometry  for the flat-sky 
  approximation is shown.}
\label{fig:1}
\end{figure}

\newpage

\begin{figure}
\figurenum{2}
\includegraphics[angle=270,width=1.1\textwidth]{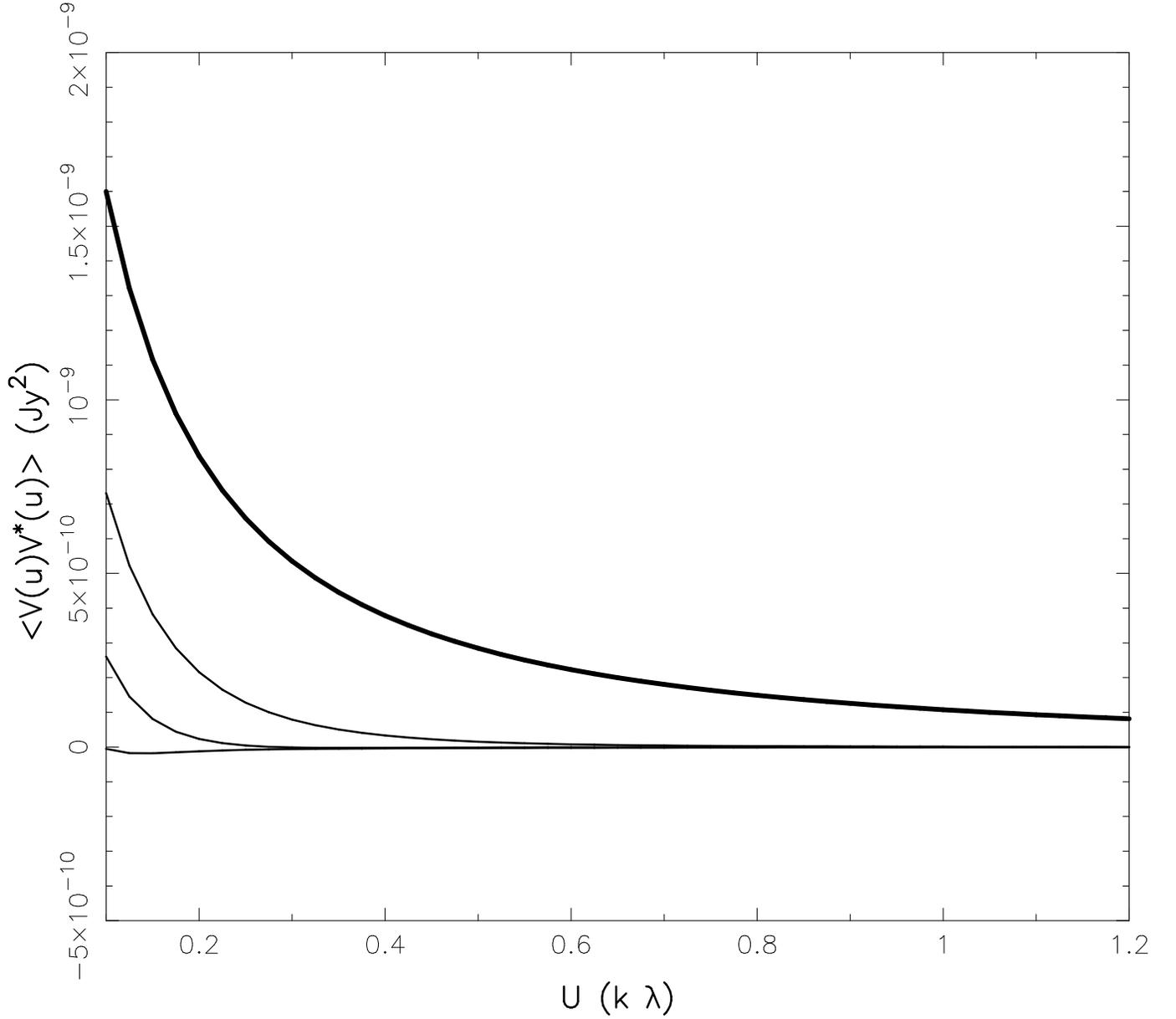}
\caption{For the central frequency $\nu = 320 \, \rm MHz$, this figure
  shows the visibility correlation function as  baseline $U$ varies.
  The thick curve shows the
  visibility correlation for $\nu = \nu'$. The other three curves
  show, from top to bottom, the visibility correlation for $\nu' - \nu
  = \{0.5, 1, 2 \} \, \rm MHz$, respectively.}
\label{fig:2}
\end{figure}

\newpage

\begin{figure}
\figurenum{3}
\includegraphics[angle=270,width=1.1\textwidth]{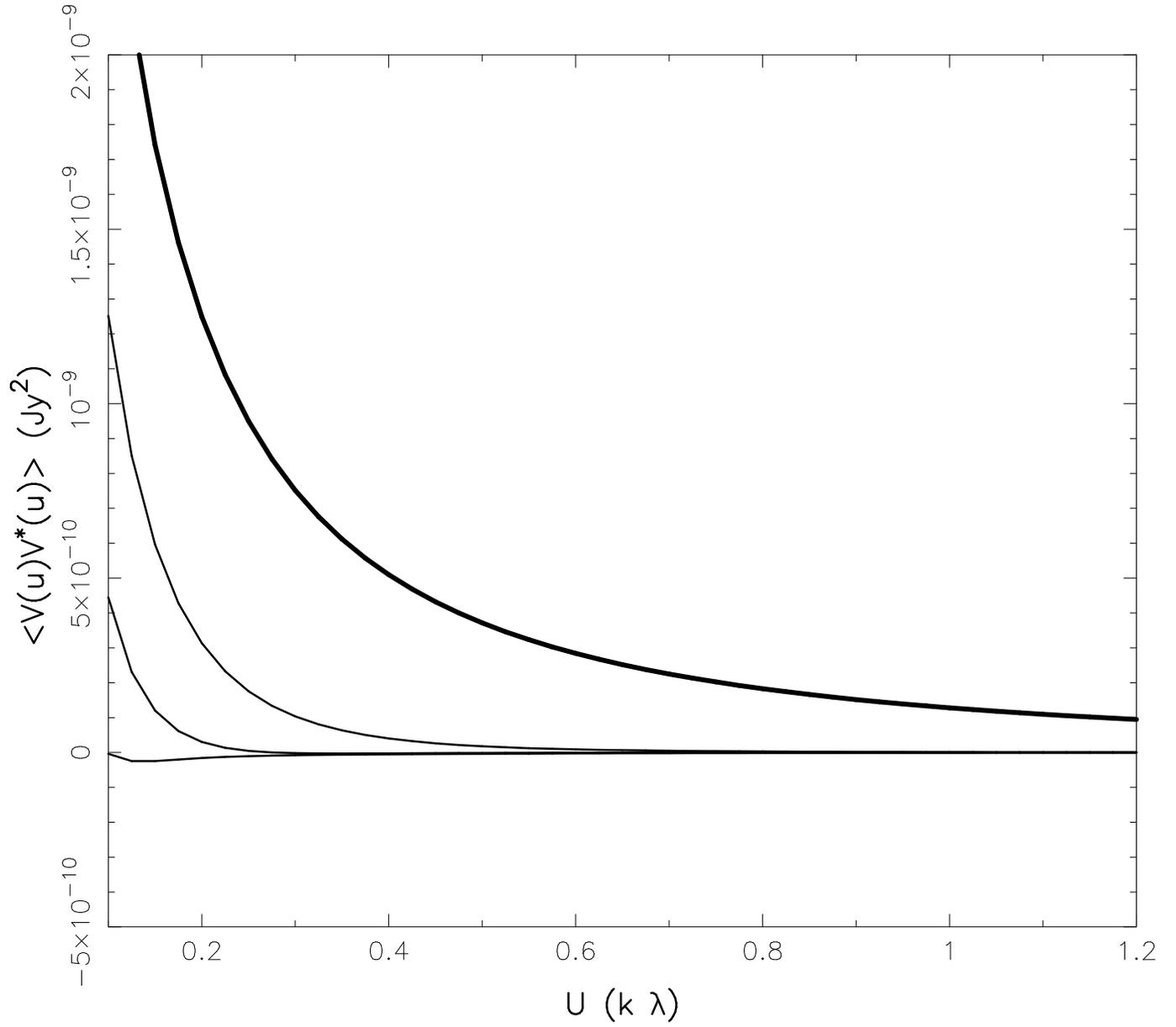}
\caption{Same as Figure~2 for the central frequency $\nu = 610 \, \rm MHz$}
\label{fig:3}
\end{figure}

\end{document}